\documentclass[preprint,12pt,authoryear]{elsarticle}

\usepackage{lipsum}
\usepackage[margin=2cm]{geometry}
\usepackage{amssymb}
\usepackage{amsmath}
\usepackage{natbib}
\usepackage{lscape}
\usepackage{multirow}
\usepackage{hhline}
\usepackage{booktabs}
\usepackage{array}
\newcolumntype{L}[1]{>{\raggedright\let\newline\\\arraybackslash\hspace{0pt}}m{#1}}
\newcolumntype{C}[1]{>{\centering\let\newline\\\arraybackslash\hspace{0pt}}m{#1}}
\newcolumntype{R}[1]{>{\raggedleft\let\newline\\\arraybackslash\hspace{0pt}}m{#1}}
\newcommand*{\thisIndent}{\hspace*{0.5cm}}
\makeatletter
\def\ps@pprintTitle{%
 \let\@oddhead\@empty
 \let\@evenhead\@empty
 \def\@oddfoot{}%
 \let\@evenfoot\@oddfoot}
\makeatother

\begin{document}

\begin{frontmatter}

\title{Enhancement of land-use change modeling using convolutional neural networks and convolutional denoising autoencoders}

\author{Guodong Du\textsuperscript{1}, Liang Yuan\textsuperscript{1}, Kong Joo Shin\textsuperscript{2}, Shunsuke Managi\textsuperscript{2}}

\address{
\textsuperscript{1}Department of Urban Engineering, Graduate School of Engineering, Kyushu University, Japan \\
\textsuperscript{2}Department of Urban Engineering, School of Engineering, Kyushu University, Japan
}

\begin{abstract}
The neighborhood effect is a key driving factor for the land-use change (LUC) process. This study applies convolutional neural networks (CNN) to capture neighborhood characteristics from satellite images and to enhance the performance of LUC modeling. We develop a hybrid CNN model (conv-net) to predict the LU transition probability by combining satellite images and geographical features. A spatial weight layer is designed to incorporate the distance-decay characteristics of neighborhood effect into conv-net. As an alternative model, we also develop a hybrid convolutional denoising autoencoder and multi-layer perceptron model (CDAE-net), which specifically learns latent representations from satellite images and denoises the image data. Finally, a DINAMICA-based cellular automata (CA) model simulates the LU pattern. The results show that the convolutional-based models improve the modeling performances compared with a model that accepts only the geographical features. Overall, conv-net outperforms CDAE-net in terms of LUC predictive performance. Nonetheless, CDAE-net performs better when the data are noisy.

\end{abstract}


\end{frontmatter}



\section{Introduction}

Land-use change (LUC) modeling is an effective approach for predicting the future land-use pattern of urban areas. LUC modeling provides supporting evidence for  urban planners and stakeholders' decision-making. Moreover, LUC modeling provides valuable information for environmental and ecological evaluation \citep{Hixson2010, Hutyra2011}. The LUC process is complicated; it is driven by natural, social and economic factors \citep{Deng2009}. To describe the process, modern LUC modeling uses a series of transitional rules, which are usually determined by assimilating several essential driving factors, such as neighborhood effect and accessibility.

Cellular automata (CA) simulates the complex transitional rules by stacking simple neighborhood rules \citep{White1993}. Given its simple but effective mechanism, CA has become the most prevalent approach in LUC modeling over the last decade \citep{Aburas2016}. CA's effectiveness also indicates the important role of neighborhood rules in LUC modeling; CA variants can enhance CA's performance by modifying, transforming or extending the mechanisms of neighborhood rule construction \citep{Sante2010, Chaudhuri2013}. 

Patch-based CA adopts a patch-based simulation strategy rather than a cell-based strategy \citep{Li2013, Chen2014, Chen2016, Li2017}. It simulates the behavior of LU patches (i.e., homogeneous cells that are spatially connected) to generate overall LU patterns, and this simulation process can be referred to as a mechanism that binds and regularizes the transitional rules of cells that are located in the same neighborhood.

Other CA variants combine the CA with statistical learning methods, in which neighborhood characteristics are usually incorporated to estimate the LU transition probability \citep[e.g.][]{Li2002, Yang2008, Al-sharif2015, du2018comparative}. In the integrated modeling system, previous studies show that the accuracy of the intermediate transition probability map greatly influences the final simulated performance \citep{CamachoOlmedo2013}. To capture precise neighborhood characteristics, \cite{Verburg2004} designed LU enrichment metrics to measure the relative abundance of LU categories in the neighborhood. \cite{Liao2016} extended the LU enrichment by assigning various distance-based influence weights. Other studies apply landscape metrics, which were originally used to analyze ecological issues, to the LUC modeling. Several typical categories of landscape metrics are used in LUC modeling studies: area metrics (e.g., largest patch index \citep{Herold2003}), shape metrics (e.g., perimeter-area ratio \citep{Chen2016}), aggregation metrics (e.g., landscape shape index \citep{Verstegen2014}, contagion \citep{Herold2003}, percentage of like adjacencies \citep{RoyChowdhury2014}), and isolation metrics (e.g., landscape similarity index \citep{Li2015}, Euclidean nearest neighbor distance \citep{Chen2016}). However, these approaches have two major limitations. First, they are limited in terms of their ability to capture complex spatial features (e.g., spatial pattern). Most metrics are designed to capture simple features such as quantity, ratio, area or edge. Moreover, the composite metrics are mainly designed to capture specific aspects of neighborhood characteristics. For instance, contagion specifically represents the aggregation/interspersion degree of neighborhood patches. Finally, these approaches derive spatial features from classified LU maps, which are relatively more homogeneous and have less spatial variance compared with the original satellite images.

Convolutional neural networks (CNN), a classic deep-learning method, may be the solution for overcoming the abovementioned limitations. CNN is well-known for its ability to process image data and extract hierarchical features \citep{Lecun2015}. It learns low-level spatial structures (e.g., edges) from its first convolutional layer and gradually stacks and extracts complex hierarchical spatial features as 'the model goes deeper'. CNN is used to solve various image processing tasks, including image classification, object detection/tracking, semantic segmentation, etc., and has been applied in various fields, including computer vision \citep[e.g.][]{Krizhevsky2012, Cox2014}, remote sensing \citep[e.g.][]{Maggiori2016, Wang2016, Long2017}, medical image analysis \citep[e.g.][]{Li2014, Qayyum2017}, etc. In particular, CNN has recently gained popularity in remote-sensing studies \citep{Nogueira2017}, which is closely related to LUC modeling studies. \cite{Makantasis2015} classified hyperspectral images using a CNN with only two layers and achieved state-of-the-art performance.

Moreover, deep learning essentially refers to multi-layered interconnected neural networks; its basic form has been used in LUC modeling since the early 2000s \citep{Li2001, Li2002}. Previous researchers have applied neural networks in various ways: standalone application \citep[e.g.][]{Liu2008, Wang2011}, integration with CA and/or other statistical methods \citep[e.g.][]{Guan2005, Grekousis2013, Li2015}, etc. These studies found that neural networks can result in reliable LU predictions. Nevertheless, other than the multi-layer perceptron (MLP), powerful neural network variants with advanced architectures are rarely used in LUC modeling studies. 

In this study, we develop an integrated modeling framework that consists of a hybrid CNN model and a DINAMICA-based CA model to simulate the LUC process of the Saitama prefecture, which is located at the western side of Japan's Greater Tokyo Area. The hybrid CNN model estimates the LU transition probability based both on spatial features learned from satellite images and on manually designed geographical features. The DINAMICA-based CA model simulates the LU pattern by referring to the generated transition probability map. We identify the improvement in predictive performance from incorporating CNN by comparing the accuracies of the transition probability maps, which are estimated using the hybrid CNN model and an MLP model that accepts only geographical features. The area under receiver operating characteristic curve (AUC-ROC) and the area under precision-recall curve (AUC-PR) are employed to evaluate the estimation accuracy. In addition, we develop a convolutional denoising autoencoder (CDAE) model, which learns latent spatial features from satellite images in an unsupervised approach, as an alternative to the supervised CNN model. This study contributes to the existing literature by 1) identifying the benefit of utilizing satellite images data through convolutional-based deep learning techniques for LUC modeling and 2) elucidating the strengths of the supervised and unsupervised approaches.

The remainder of this paper is organized as follows. Section 2 describes the CNN, CDAE and CA models and the performance evaluation metrics. Section 3 describes the study area and data. Section 4 presents the results of the model evaluations and the simulated LU maps. Section 5 discusses the effects of convolutional filter size, spatial weight layer and pooling and provides visualizations for the two convolutional-based models. Section 6 concludes.

\section{Methodology}

\subsection{Neural network models}

\subsubsection{Geo-net}
As the reference model, we develop a neural network model, which includes a set of conventional geographical feature and excludes features linked to satellite images. This model is compared with the hybrid CNN model, which includes both geographical features and features linked to satellite images.
Specifically, the reference model is an MLP with ReLU as the non-linear activation function and Sigmoid as the classifier. 
We construct and use the commonly adopted geographical features, which are land-use enrichment \citep{Verburg2004}, proximity factors, land price, population density and physical factors, covering the neighborhood characteristics, accessibility, socio-economical and physical factors for the cell of interest. Table 1 describes the geographical features.

\begin{table}[h]
\small
\caption{Geographical features used in LUC models}
\begin{tabular}{p{5cm} p{12.5cm}}
\toprule
Category & Description  \\
\midrule
\textit{LU enrichment} & \multirow{4}{12.5cm}{The relative abundance of certain LU category in the neighborhood calculated by \(\displaystyle Enrichment_{i,k,d} = \frac{n_{k,d,i} / n_{d,i}}{N_k/N}\), where $Enrichment_{i,k,d}$ is the enrichment of neighborhood $d$ of location $i$ with land use type $k$, $n$ denotes the number of cells in the neighborhood, and $N$ denotes the number of cells in the whole map} \\
\thisIndent LU enrichment of forest & \\
\thisIndent LU enrichment of agri. & \\
\thisIndent LU enrichment of built-up & \\
\thisIndent LU enrichment of water body & \\
\textit{Accessibility} & \\
\thisIndent Distance to major roads & \multirow{3}{12.5cm}{The nearest Euclidean distance from the given cell to certain geographical objects} \\
\thisIndent Distance to railway stations & \\
\thisIndent Distance to urban center & \\
\textit{Socio-economical factors} & \\
\thisIndent Land price & The estimated land price of given cell; the raw data is provided by the Ministry of Infrastructure, Land and Tourism of Japan \\
\thisIndent Population density & The population density of given cell, provided by the Ministry of Infrastructure, Land and Tourism of Japan \\
\textit{Physical factors} & \\
\thisIndent Elevation & The elevation of given cell, provided by SRTM (Shuttle Radar Topographic Mission) database \\
\thisIndent Coordinates & The coordinates of given cell \\                                                                                                                             
\bottomrule
\end{tabular}
\footnotesize\flushleft{
\textit{Notes:}\\
The land price map is interpolated from a polypoint land price map, in which each record is collected by field survey, by using ordinary Kriging interpolation method.}
\end{table}

\subsubsection{Conv-net}
CNN is a special class of neural networks that uses convolutional operations in place of matrix multiplication in the hidden layers.
A typical hidden CNN layer consist of three parts: 1) a convolutional layer that performs several parallel convolutions, 2) non-linear activation layer, 3) pooling layer that replaces the output of the net at a certain location with a summary statistical of the nearby outputs.
Compared with the conventional fully connected networks, CNN can be regarded as a locally connected network, which allows each hidden unit to connect to a small subset of the input units. Specifically, when processing images, each hidden unit will connect only to a small contiguous region of pixels in the input. 
This architecture grants CNN higher computational efficiency and the ability to capture the local pattern. The pooling operation further grants CNN an invariance in small local translation, which is a particularly useful property for tasks when identifying whether the existence of some features is more important than their location (e.g., object detection). 

Various meta-architectures of CNN have been developed in these years. 
Although their performances have been evaluated using ImageNet classification or similar classification tasks, their suitabilities to the LUC modeling problem still require examination, due to LUC modeling's distinct characteristics. Specifically, 1) ordinary satellite images, such as Landsat images, have relatively low resolution, vague edges and barely distinguishable objects; 2) local satellite image patches may contain redundant information because of the high spatial autocorrelation; and 3) the LUC modeling's desired features may differ. 

In this study, we build the CNN architecture using as reference the designs of three classic meta-architectures: Alex-net \citep{Krizhevsky2012}, which has a relatively large convolutional filter size; VGG \citep{Simonyan2014}, which has a small kernel size but a deep architecture; and ResNet \citep{He2016}, which constructs residual blocks to facilitate a better gradient flow. 
The architectures are finally determined according to the results of trial-and-error experiments.
Table 2 presents the model architectures for modeling three LU transitions.
These architectures include several noteworthy aspects: 1) we use a stride of 3 rather than 2 while downsampling, given the input satellite image that represents the neighborhood usually has an odd-numbered size number to guarantee that the cell of interest is centrally located; 2) although the three models' architectures are determined independently, their convolutional architectures turn out to be identical, which may indicate that CNN serves the same function even when the target transitional rules are different; 3) the pooling layers are used only next to the first two convolutional layers because the posing information, such as the direction to the center, may become important at a higher hierarchical level; and 4) the number of filters in our study is large for a binary classification task, possibly because of the high variance of spatial patterns.

\begin{table}[h]
\centering
\small
\caption{Architectures of the conv-nets}
\begin{tabular}{|c|c|c|}
\hline
Forest to agri. & Agri. to forest & Agri. to built-up \\ \hline
conv-128        & conv-128        & conv-256          \\ \hline
\multicolumn{3}{|c|}{spatial weight}                    \\ \hline
\multicolumn{3}{|c|}{max pooling}                       \\ \hline
conv-256        & conv-256        & conv-512          \\ \hline
\multicolumn{3}{|c|}{spatial weight}                    \\ \hline
\multicolumn{3}{|c|}{max pooling}                       \\ \hline
conv-512        & conv-512        & conv-1024         \\ \hline
\multicolumn{3}{|c|}{spatial weight}                    \\ \hline
conv-1024       & conv-1024       & conv-2048         \\ \hline
\multicolumn{3}{|c|}{spatial weight}                    \\ \hline
\multicolumn{3}{|c|}{global average pooling}            \\ \hline
dense-1036      & dense-1036      & dense-2048        \\ 
dense-400       & dense-400      & dense-800         \\ 
dense-80        & dense-80      & dense-300        \\ 
dense-7         & dense-7                & dense-120          \\ 
 & & dense-60 \\ \hline
\multicolumn{3}{|c|}{Sigmoid}  \\ \hline             
\end{tabular}
\footnotesize\flushleft{
\textit{Notes:}\\
1. Conv-N denotes a convolutional block composing of a convolutional layer with $3\times3$ filter size and N of filters, a Batchnorm layer and a ReLU layer. \\
2. Max pooling layer has a kernel size of $3\times3$ and a stride of 3. \\
3. Dense-N denotes fully connected (dense) layer with N of hidden units. \\
4. Input image tensor has the shape of ($7\times26\times26$). 
}
\end{table}

According to the neighborhood effect, the influence of a neighboring cell decreases with its distance to the central cell. To introduce this mechanism into conv-net, we specifically design a regularization layer named the spatial-weight layer. It is described as
\begin{align}
y &= x * SW \\
SW_{i, j, d} &= e^{a_d dist_{i, j}} + b_d
\end{align}
$y$ and $x$ are the output and input, respectively. $SW$ is the spatial weight. $i$, $j$, $d$ denote the location in an image tensor, $dist_{i, j}$ is the Euclidean distance from location(i, j) to the center, $a_d$ and $b_d$ are the trainable parameters. This spatial weight layer can be regarded as imposing a distance-based prior on the spatial feature map, which exponentially reduces the influence of spatial features that are far from the center. 

The CNN model is placed parallel to an MLP model that accepts the geographical features as input. Both networks are then connected to an MLP classifier composed of several fully connected hidden layers (see Figure 1). This hybrid CNN model is trained as a whole. 

\begin{figure}[h]
\centering
\includegraphics[width=17cm]{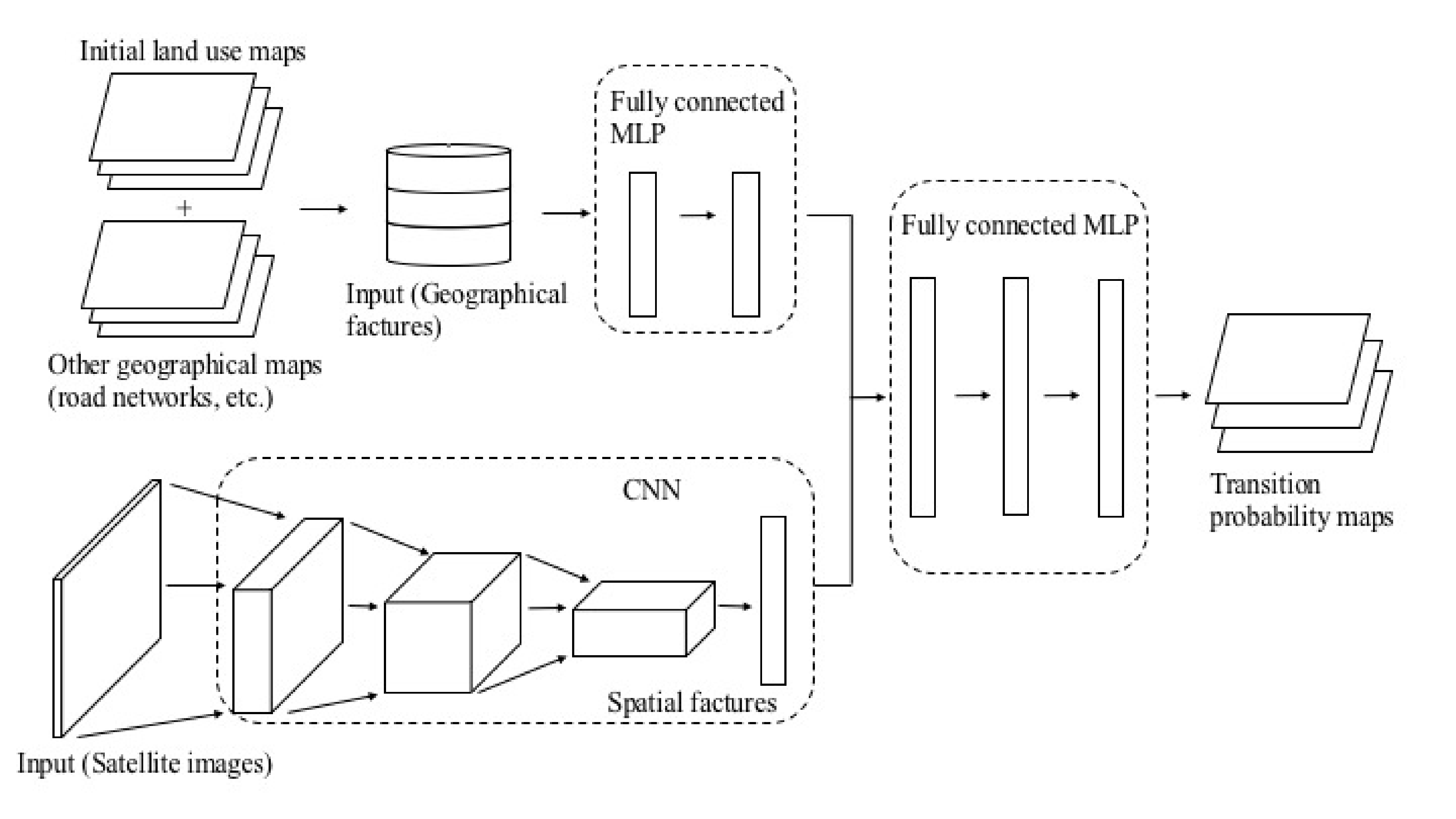}
\caption{The structure of conv-net}
\footnotesize\flushleft{
}
\end{figure}

\subsubsection{CDAE-net}
An autoencoder (AE) is an unsupervised learning algorithm that copies the input to the output. It is essentially a neural network; thus, it can be trained by backpropagation. An AE consists of two parts: the encoder and the decoder. The encoder maps the input into hidden representations, and the decoder reconstructs the input from the hidden representations. The general form of an AE is
\begin{align}
h &= s(Wx + b) \\
y &= s(W'h + b')
\end{align}
$x$ is the input; $h$ is the latent representation; $y$ is the output; $s$ is non-linear activation function such as Sigmoid; $W$ and $W'$ are the encoder and decoder weights, respectively; and $b$ and $b'$ are the encoder and decoder biases, respectively. AE usually has an undercomplete architecture, in which the dimension of $h$ is smaller than $x$, to extract useful features from input rather than just learn an identity function. 

A denoising autoencoder (DAE) \citep{Vincent2008} is a variant of AE that is designed to capture more robust features by reconstructing the input from a corrupted version of it. The general form of a DAE is
\begin{align}
h &= s(W\tilde{x} + b) \\
y &= s(W'h + b')
\end{align}
$\tilde{x}$ is a copy of x that has been corrupted by some form of noise. The noise injection forces the DAE to capture the statistical dependencies between the inputs by causing the DAE to undo the effect of the corruption process. 

In this study, we incorporate the convolutional operation into a DAE and develop a convolutional denoising autoencoder (CDAE) model to tackle the possible data problems that cannot be effectively addressed by CNN, namely, the redundant spatial information and the satellite image noise. The CDAE model is loosely combined with an MLP classifier to estimate the transition probability. Specifically, original input from satellite images is fed into the CDAE model to produce the latent representation, and the latent representation is then fed into an MLP classifier together with geographical features. 

Figure 2 illustrates the architecture of this CDAE-net, and Table 3 presents the architectures of its encoder and classifier. For each convolutional or pooling layer in the encoder, the decoder has a deconvolutional layer (transposed convolutional layer) or an unpooling layer with the same configuration at the corresponding location, and the encoder and decoder weights are untied. Although DAE does not need to be undercomplete, we use a 'bottleneck' hidden layer with 567 hidden units, to extract the most salient features.

\begin{figure}[h]
\centering
\includegraphics[width=17cm]{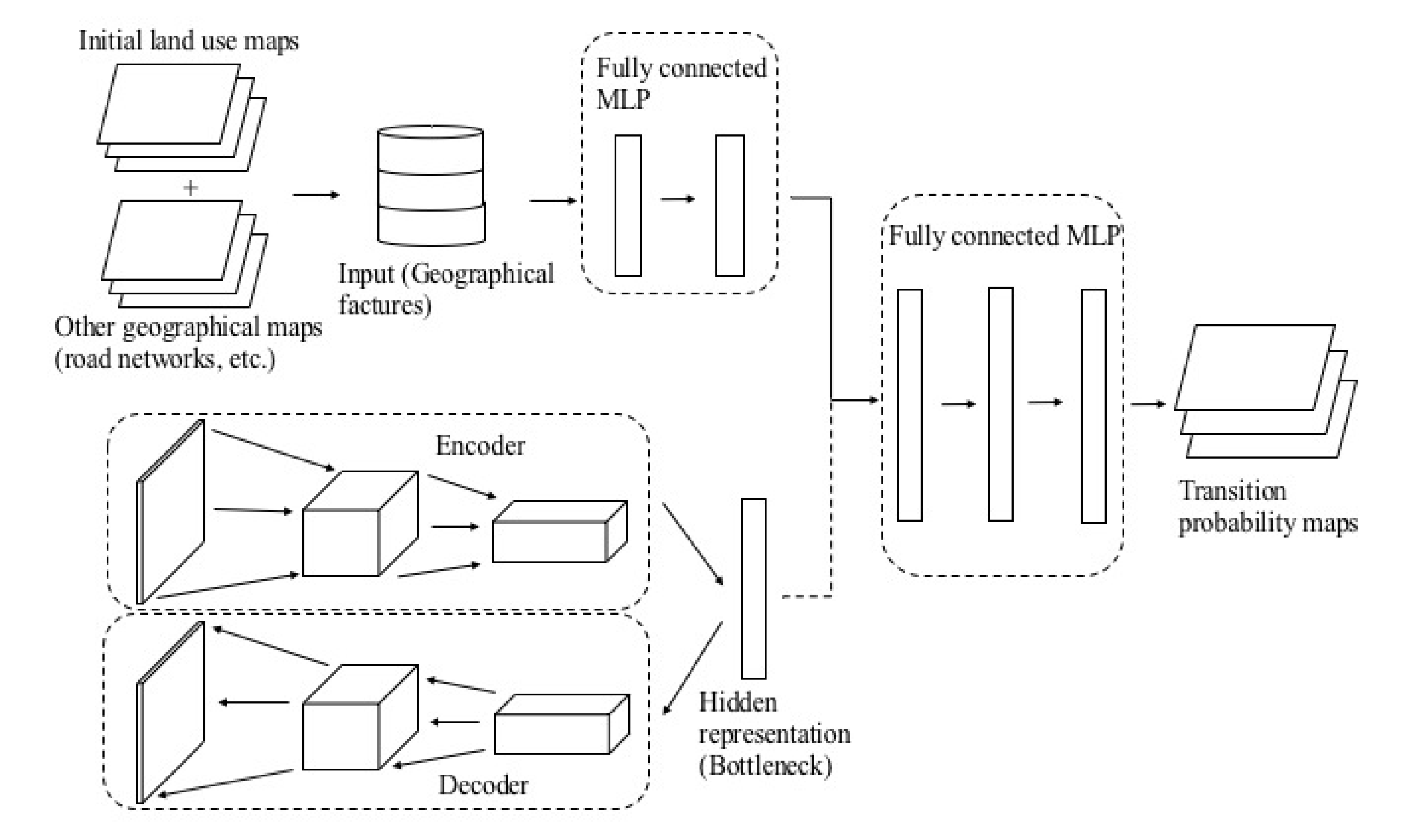}
\caption{The structure of cdae-net}
\end{figure}

\begin{table}[h]
\small
\centering
\caption{Architectures of the CDAE-nets}
\begin{tabular}{|c|c|c|}
\hline
Forest to agri. & Agri. to forest & Agri. to built-up \\ \hline
\multicolumn{3}{|c|}{conv-64}                    \\ \hline
\multicolumn{3}{|c|}{max pooling}                       \\ \hline
\multicolumn{3}{|c|}{conv-128}                    \\ \hline
\multicolumn{3}{|c|}{max pooling}                       \\ \hline
dense-579      & dense-579      & dense-579        \\ 
dense-579      & dense-579      & dense-579        \\ 
dense-579      & dense-579      & dense-579        \\ 
 & & dense-100 \\ \hline
\multicolumn{3}{|c|}{Sigmoid}  \\ \hline             
\end{tabular}
\footnotesize\flushleft{
\textit{Notes:}\\
1. Conv-N denotes a convolutional block composing of a convolutional layer with $3\times3$ filter size and N of filters, a Batchnorm layer and a ReLU layer. \\
2. Max pooling layer has a kernel size of $3\times3$ and a stride of 3. \\
3. Dense-N denotes fully connected (dense) layer with N of hidden units. \\
4. Input image tensor has the shape of ($7\times81\times81$). \\
5. For each convolutional layer in the encoder, there is a corresponding deconvolutional layer in the decoder; for each each max pooling layer, there is a corresponding unpooling layer in the decoder.
}
\end{table}

CDAE-net and conv-net differ in two major ways in terms of the architecture: 1) CDAE-net and classifier are separately trained independently; 2) CDAE-net have no global pooling layer at the end of its architecture, and its output 3D tensor is raveled and directly fed into the fully connected classifier. 

\subsubsection{Model training}
As the main regularization method, we add Gaussian noise to the gradient \citep{neelakantan2015adding}, which is demonstrated effective for training deep networks, and discard dropout because our experiments show that Batchnorm eliminates its need.
We use the image jitter method proposed by \cite{Krizhevsky2012} to produce training samples with varying degrees of illumination to improve the model's generalization performance; however, we do not use image flip to preserve the pose information.
We use binary cross-entropy as the loss function for training classifier and mean square error as the loss function for training CDAE.
We use stochastic gradient descend as the optimization method, and use the parameter initialization method suggested by \cite{Glorot2010}, which shortens the convergence time by approximately 0.8 compared with the initialization method based on random sampling from Gaussian distribution.
For each neural network model, we perform fine-tuning on a set of hyperparameters, including the learning rate, the learning rate decay frequency, momentum, and the Gaussian noise coefficient. 

\subsection{Cellular automata}

The CA model is a variant of DINAMICA. DINAMICA defines two main vicinity-based transitional functions, expander and patcher, to simulate land-use patch dynamics in a stochastic multi-step approach \citep{Soares-Filho2002}. The expander function is dedicated to the expansion or contraction of the previous patches of a certain land-use class, and the patcher function is designed to generate new patches. The two processes are merged using the following calculation:
\begin{equation} 
Q_{ij} = r \times expander + s \times patcher
\end{equation}
where $Q_{ij}$ is the total number of transitions from land-use class $i$ to $j$; $r$ and $s$ are the percentages performed by the expander and patcher functions, respectively; and $r$ + $s$ = 1. 

The expander function is defined as follows:
\begin{align}
& if \quad n_j > 3 \quad or \quad P(ij)(xy) > t \quad then \quad P'(ij)(xy) = P(ij)(xy) \\
& else \quad P'(ij)(xy) = p(ij)(xy) \times \sqrt{\frac{n_j}{4}}
\end{align}
$P(ij)(xy)$ denotes the transition probability from land-use class $i$ to $j$, t denotes a preset threshold, and $n_j$ denotes the number of cells of land-use class $j$ occurring in a 3$\times$3 window.
Both processes use a stochastic selection mechanism to select seeds (the center cell in a transition patch). This mechanism selects the seeds while prioritizing high over low transition probabilities with a certain degree of randomness. The patch size is drawn from a lognormal distribution, and the patch shape or compactness is determined by a parameter called isometry.

The transition quantity is determined based on transition probability maps. Specifically, we manually set a threshold for transition probability, and the quantity of cells that have transition probability above the threshold is used as the total quantity for CA simulation.

\subsection{Evaluation metrics}

We use AUC-ROC and AUC-PR to assess the predictive performances of the neural network models. AUC-ROC has been widely used as a quantitative measure for assessing classification performance. Specifically, an AUC-ROC value of 0.5 is the random baseline, and values below 0.5 indicate a systematically incorrect model \citep{Jansen2012}. 
ROC is frequently used to evaluate the quality of transition probability \citep{Pontius2011}. 
However, ROC has two limitations: (1) it provides an overall evaluation on the performances of both negative and positive labels; and (2) it presents an overly optimistic view of an algorithm's performance if the data are highly imbalanced \citep{Davis2006}.
AUC-PR can be used as a complement to AUC-ROC. AUC-PR is a sequence of precision and recall values with varying thresholds. It provides a more specific assessment of a model's ability to predict changed areas and exclude the influence of data imbalance.

We use overall accuracy, quantity/allocation disagreement, Cohen's Kappa statistic, Kappa simulation and fuzzy Kappa simulation to evaluate the agreement between a simulated LU map and an actual LU map.
Quantity/allocation disagreement and Kappa statistic are commonly used cell-to-cell evaluation metrics. Quantity disagreement is defined as the difference between two maps in terms of the quantity of land-use category mismatch. Allocation disagreement is the difference between two maps in terms of the mismatch of each land-use category's spatial allocation. Kappa statistic is a classic map comparison method, which excludes the proportion of agreement by chance.
Nevertheless, the two metrics cannot eliminate the influence of LU persistence, which may lead to over-estimation of the predictive performance of LUC model. In order to deal with this limitation, we employ Kappa simulation and fuzzy Kappa simulation. Both metrics exclude the influence of LU persistence by incorporating the initial land use map, but Kappa simulation is a cell-to-cell metrics, while fuzzy Kappa simulation is a vicinity-based metrics, which introduces fuzzy set theory to account for the LU category similarity and neighborhood similarity. We only consider the neighborhood similarity in this study given the limited number of LU categories, and use a Gaussian distance decay function to specify the agreement level with respect to distance within a neighborhood. \cite{Taylor2003} and \cite{VanVliet2013} provides demonstrations and technical details of fuzzy Kappa and fuzzy Kappa simulation, respectively.

\section{Implementation}

Our study area is the Saitama prefecture of Japan, which is located at the western side of the Greater Tokyo area. It covers an area of 3,798 $km^2$ and has a population size of 7,237 thousand. Most parts of Saitama can be regarded as Tokyo suburbs, and Saitama's urban area is constantly expanding due to immigration to the Greater Tokyo Area.

We collect Global Land Survey (GLS) satellite image datasets for 2000, 2005 and 2010. GLS datasets have eight bands, within which band 8 (Panchromatic) has a resolution of 15 meters, and the other seven bands have resolutions of 30 meters each. All bands except band 8 are resampled into 15 meters and combined for LU classification. We classify four LU categories (water, agriculture, forest and built-up) using the supervised classification algorithm provided in ERDAS IMAGINE V2016 (Hexgon Geospatial, U.S.). Table 4 presents the confusion matrices of LUC, and Figure 3 shows LU maps of the three years. We exclude the LU transitions that have transition rates below 1\% and are left with transition from forest to agriculture, transition from agriculture to forest, and transition from agriculture to built-up as the modeling objects. 

\begin{table}[h]
\small
\centering
\caption{Confusion matrices from 2000 to 2005 and from 2005 to 2010}
\begin{tabular}{p{.7cm} lllll}
\toprule
 & & \multicolumn{4}{c}{2005} \\
 &  & Built-up          & Forest            & Agri.             & Water body       \\
\hhline{~~----}
\multirow{5}{.7cm}{2000}  & Built-up     & 4090795 (98.99\%) & 13672 (0.33\%)    & 20508 (0.50\%)    & 7522 (0.18\%)    \\
 & Forest       & 22787 (0.72\%)    & 2693884 (85.10\%) & 435253 (13.75\%)  & 13672 (0.43\%)   \\
 & Agri.        & 419344 (7.48\%)   & 575959 (10.27\%)  & 4601687 (82.06\%) & 10937 (0.20\%)   \\
 & Water body   & 478 (0.39\%)      & 888 (0.72\%)      & 683 (0.55\%)      & 121397 (98.34\%) \\
\midrule
 & & \multicolumn{4}{c}{2010} \\
 & & Built-up          & Forest            & Agri.             & Water body       \\
\hhline{~~----}
\multirow{5}{.7cm}{2005} & Built-up     & 4601286 (98.30\%) & 33151 (0.71\%)    & 37887 (0.81\%)    & 8530 (0.18\%)    \\
 & Forest       & 20894 (0.66\%)    & 2564788 (80.75\%) & 565762 (17.81\%)  & 24863 (0.78\%)   \\
 & Agri.        & 605314 (11.99\%)  & 345704 (6.85\%)   & 4085922 (80.93\%) & 11839 (0.23\%)   \\
 & Water body   & 947 (0.77\%)      & 591 (0.48\%)      & 591 (0.48\%)      & 121397 (98.28\%) \\
\bottomrule
\end{tabular}
\footnotesize\flushleft{
\textit{Notes:}\\
1. The confusion matrix is presented as num of cells (the percentage). \\
2. The percentage is calculated by $(num_{t} - num_{t-1}) / num_{t-1}$ where $num$ denotes the number of cells and $t$ denotes the time.}
\end{table}

\begin{figure}[h]
\centering
\includegraphics[width=17cm]{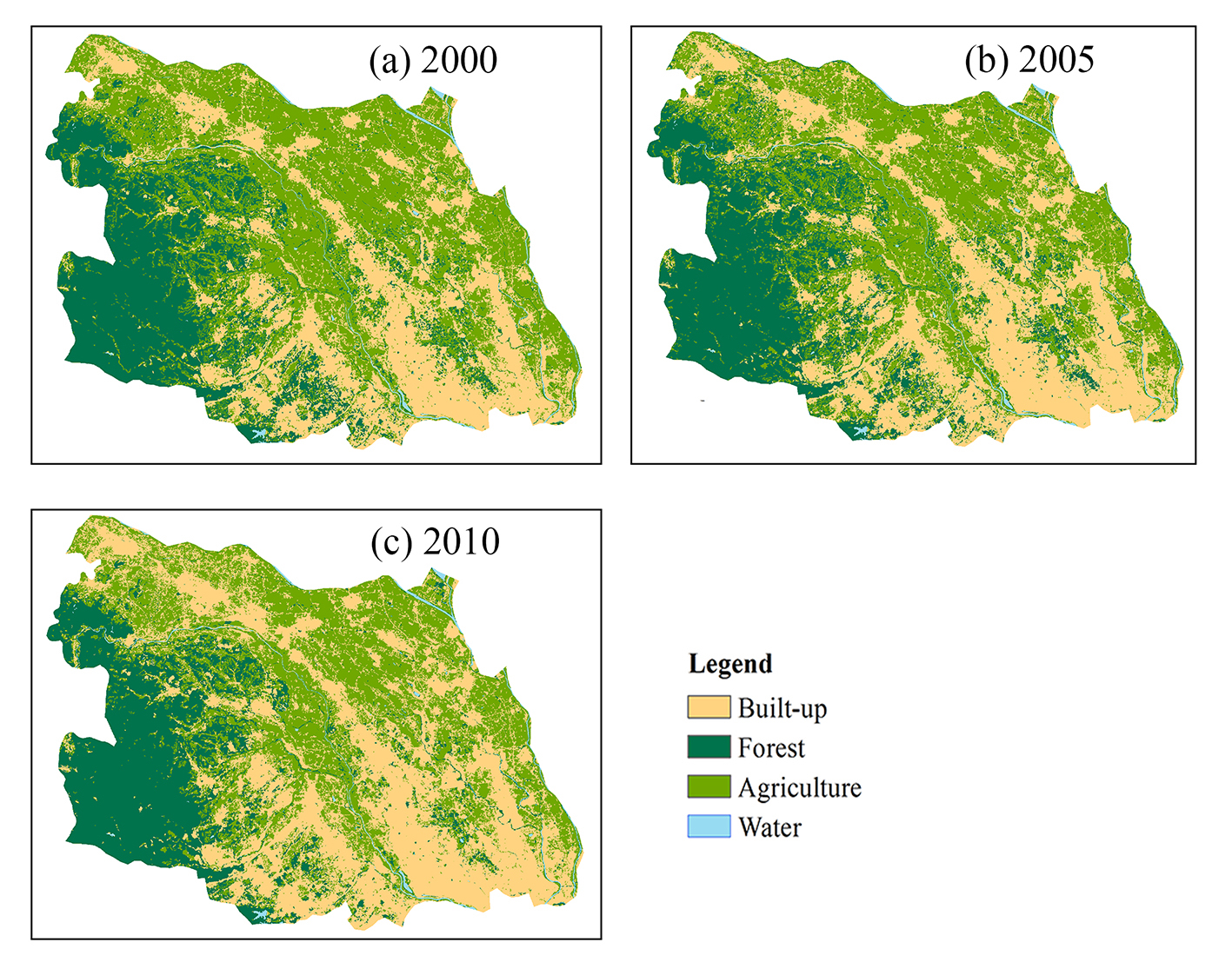}
\caption{The actual LU maps in Saitama prefecture of Greater Tokyo Area for 2000, 2005 and 2010}
\end{figure}

Given the possibility of varying transitional rules being present among the three LU transition types, we separately develop transition probability estimation models and CA models for each of them. Geographic features are derived based on spatial data collected from the Ministry of Infrastructure, Land and Tourism of Japan, except for LU enrichment, which is directly calculated from LU maps. Satellite image patches of a certain size are cropped from the satellite images and used as input for the conv-net or CDAE model. The input image size is determined based on previous evidence on the effect of neighborhood size and trial-and-error experiments. The image input size is $27\times27$ for conv-net and $81\times81$ for CDAE-net. In addition, the satellite input image has seven bands, excluding band 6 (thermal) due to its low spatial variation.

The LUC models are trained on the 2000 and 2005 datasets, validated on a subset of data for 2005 and 2010, and tested on the whole dataset for 2005 and 2010. To minimize the spatial autocorrelation between the validation set and the test set to facilitate an unbiased model evaluation, the validation set is extracted from a sub-region of Saitama covering approximately 15\% of the total area.
In terms of sampling, previous studies have normally used random or stratified sampling to avoid the influence of spatial autocorrelation. However, the mini-batch learning criterion of a neural network naturally mitigates the influence of spatial autocorrelation to some extent. We use a bootstrap over-sampling strategy in this study. Specifically, a mini-batch of data is randomly sampled from the dataset with replacement, and samples belonging to negative and positive labels have the same proportion within a mini-batch. Over-sampling could make the model prone to over-fitting. To address this issue, we consider the fine-tuning of the Gaussian noise coefficient. More discussion on the over-sampling is provided by \cite{Batista2004}.

\section{Results}

\subsection{Evaluation on the modeling performances}

Table 5 summarizes the evaluation results for geo-net, conv-net, CDAE-net. 
AUC-ROC and AUC-PR are calculated based on the transition probability estimated by the three models.
For modeling all three LU transitions, conv-net and CDAE-net consistently outperform the geo-nets for all the evaluation metrics. Particularly, conv-net and CDAE-net outperform the geo-nets by approximately 0.02$\sim$0.10 in terms of AUC and approximately 0.053$\sim$0.15 in terms of the Kappa statistic. These results confirm that the use of convolutional-based neural networks to extract spatial features from satellite images improves LUC modeling performance.

\begin{table}[h]
\small
\centering
\caption{Performance evaluation of the estimated transition probabilities}
\begin{tabular}{p{3cm} lll}
\toprule
LU transition & Model & AUC-ROC           & AUC-PR      \\
\midrule
\multirow{3}{3cm}{Forest to agri.}   & Geo-net  & 0.886 & 0.595 \\
                  & Conv-net & 0.944 & 0.714 \\
                  & CDAE-net & 0.912 & 0.675 \\
                  \hhline{~---}
\multirow{3}{3cm}{Agri. to forest}   & Geo-net  & 0.863 & 0.397 \\
                  & Conv-net & 0.905 & 0.493 \\
                  & CDAE-net & 0.880 & 0.415 \\
                 \hhline{~---}
\multirow{3}{3cm}{Agri. to built-up} & Geo-net  & 0.659 & 0.215 \\
                  & Conv-net & 0.694 & 0.239 \\
                  & CDAE-net & 0.714 & 0.268 \\
\bottomrule
\end{tabular}
\end{table}

The results indicate the validity of all the three models to learn pattern from the spatial data and to determine the transition rules, given that the AUC-ROCs of the three models are much larger than 0.7.
However, the results exhibit substantially different predictive performances for the three LU transitions. The AUC-PR obtained from the LUC models for transitions from forest to agriculture, agriculture to forest, and agriculture to built-up are 0.69, 0.42 and 0.26, respectively; the highest AUC-ROC for the three transitions are 0.92, 0.89, 0.72, respectively.
According to these results, the modeling performance for the transition from agriculture to built-up is significantly worse than the other two transitions.

Two possible reasons explain the lower performance of agriculture to built-up. First, in the LU classification, agriculture and built-up are relatively difficult to differentiate because they usually have interspersing spatial distributions, particularly in the suburbs and frequently appear as similar colors on the satellite images. 
According to the results of the LU classification assessment provided by the supervised classification algorithm, the classification accuracies of agriculture and built-up (three-years average of 85
\% and 83\%, respectively) are significantly lower than forest (three-years average of 89\%).
Therefore, the LUC modeling of the transition from agriculture to built-up may suffer more from the data noise problem than the other two transitions, which then leads to the relatively poor performance.
Second, the lower transition performance may be driven by the lack of information regarding individual decision-making regarding the transitions from agriculture to built-up in study area. 
Given that the Saitama area has no intensive urban development plan built-up and agriculture are commonly privately owned, the individual decision-making factor would act as an important determinant of the LUC process of agriculture to built-up. 
However, individual decision-making cannot be captured through spatial features, including satellite images; this information needs to be collected separately \citep{du2018comparative}.

The results indicate that, due to the capability of data denoising, CDAE-net outperforms conv-net when the image data is noisy. For the transition from agriculture to built-up, CDAE-net outperforms conv-net by approximately 0.03 in terms of AUC-PR.
However, conv-net outperforms CDAE-net when data quality is better. 
For transitions between agriculture and forest, conv-net outperforms CDAE-net by approximately 0.01$\sim$0.02 in terms of AUC-PR; this result may be induced by CDAE-net's data compression process, which inevitably discards some useful spatial information. This compression process would be preferred when data are noisy but it may lower the performance otherwise.

\subsection{Land-use simulations}

We simulate the LU maps for 2010 by using the DINAMICA-based CA based on the transition probability maps produced by the geo-net, conv-net and CDAE-net, respectively. In addition, we simulate  LU map based on the transition probability of agriculture to forest and forest to agriculture predicted by the conv-net and the transition probability of agriculture to built-up predicted by the CDAE-net. 
Figure 4 shows the comparison between simulated and actual LU maps for 2010, and Table 6 presents the evaluation metrics of the simulated maps, including accuracy, quantity/allocation disagreement, Kappa statistic, Kappa simulation and fuzzy Kappa simulation with neighborhood sizes of $3\times3$, $7\times7$ and $11\times11$.

\begin{figure}
\centering
\includegraphics[width=16cm]{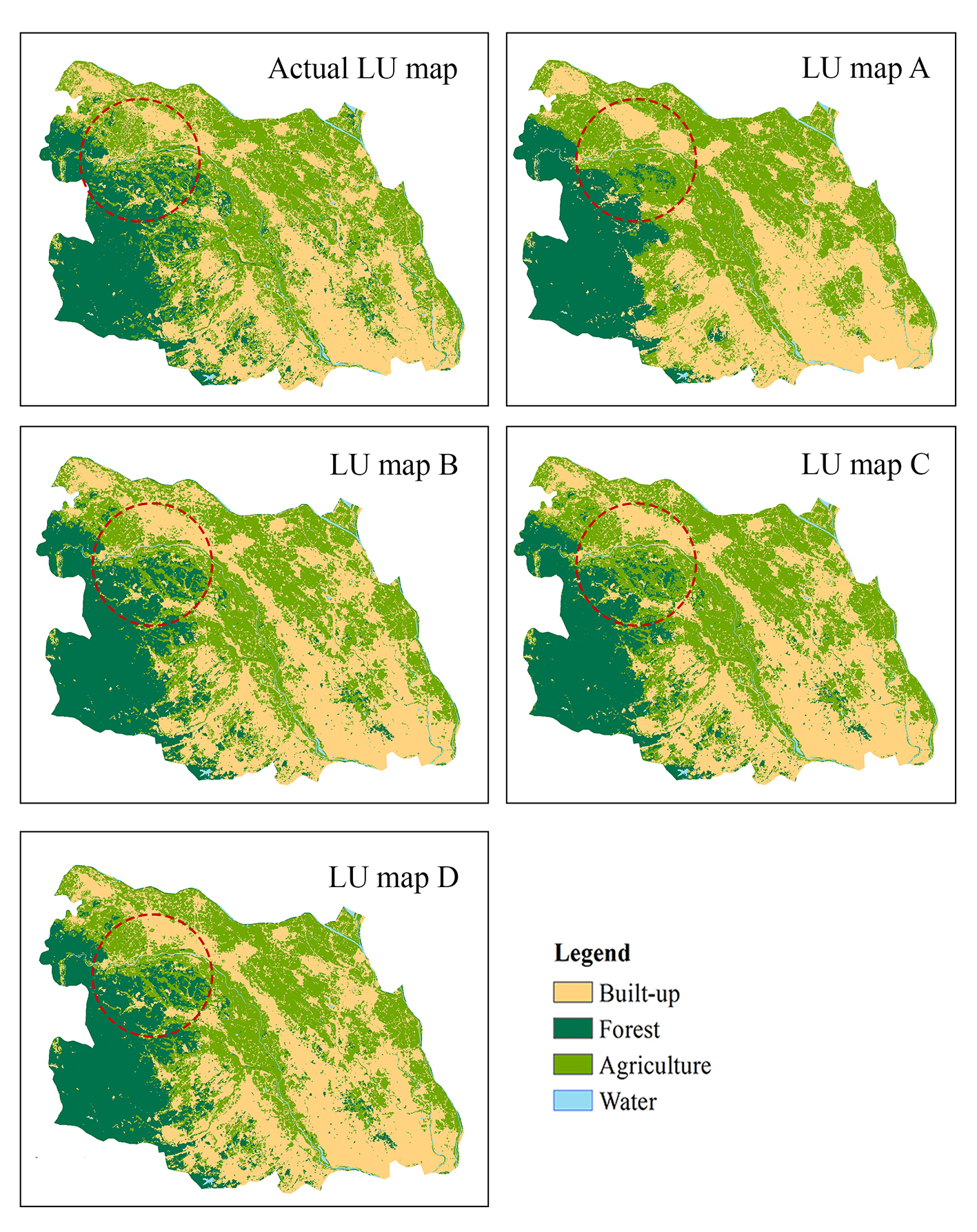}
\caption{The actual and simulated LU maps for 2010}
\begin{minipage}{\linewidth}
\footnotesize
\emph{\\Notes:\\}
1. Land use maps of A, B and C are simulated by using CA based on the transition probability predicted by the geo-net, conv-net and CDAE-net, respectively. \\
2. Land use map of D are simulated based on the transition probability of agriculture to forest and forest to agriculture predicted by the conv-net and the transition probability of agriculture to built-up predicted by the CDAE-net. 
\end{minipage}
\end{figure}

\begin{table}
\small
\centering
\caption{Performance evaluation of the simulated LU maps}
\begin{tabular}{p{1cm} p{1.6cm} p{1.6cm} p{1.6cm} p{1.6cm} p{1.6cm} p{1.6cm} p{1.6cm} p{1.6cm}}
\toprule
\multirow{2}{1.7cm}{LU map} & \multirow{2}{1.7cm}{Accuracy} & \multicolumn{2}{c}{Disagreement}  & \multirow{2}{1.7cm}{Kappa statistic} &  \multirow{2}{1.7cm}{Kappa simulation} & \multicolumn{3}{c}{Fuzzy Kappa simulation}    \\
\hhline{~~--~~---}
         &                       &     Quantity                   &    Allocation             &                 &                         & $3\times3$ & $7\times7$ & $11\times11$ \\
\midrule
A        & 0.897                 & 0.016                   & 0.086           & 0.812            & 0.337                  & 0.338      & 0.340      & 0.345        \\
B        & 0.908                 & 0.011                   & 0.081           & 0.830             & 0.391                  & 0.414      & 0.455      & 0.484        \\
C        & 0.901                 & 0.013                   & 0.078           & 0.822            & 0.376                  & 0.399      & 0.437      & 0.455        \\
D        & 0.914                 & 0.005                   & 0.080           & 0.842            & 0.412                  & 0.435      & 0.474      & 0.501       \\
\bottomrule
\end{tabular}
\footnotesize\flushleft{
\textit{Notes:}\\
1. Land use maps of A, B and C are simulated by using CA based on the transition probability predicted by the geo-net, conv-net and CDAE-net, respectively. \\
2. Land use map of D are simulated based on the transition probability of agriculture to forest and forest to agriculture predicted by the conv-net and the transition probability of agriculture to built-up predicted by the CDAE-net. 
}
\end{table}

The order of the performances of the three LUC models is conv-net + CA $>$ CDAE-net + CA $>$ geo-net + CA; this result is consistent with the difference of transition probability prediction performances.
The LU map D in Table 6 combines the transition probability predictions with the highest accuracy, which also exhibits the highest simulation performance.
The values of Kappa simulation and fuzzy Kappa simulation are significantly lower than the values of Kappa statistic; this result is plausible given the exclusion of the influence of LU persistence. The significant difference between the values of Kappa statistic and Kappa simulation indicates that the Kappa simulation is stricter metrics than the Kappa statistic. 
In terms of the fuzzy Kappa simulation, the value of fuzzy Kappa simulation obtained from the geo-net + CA increases much less with the increase of neighborhood size when compared with the increase of values obtained from the conv-net + CA or the CDAE-net + CA.
Compared with Kappa simulation, fuzzy Kappa simulation provides additional evaluation aspect of LUC model's capability to yield the 'near hits' (i.e. the LUC model does not precisely allocate the LU transition to the target cell but allocates the LU transition to the cells within the neighborhood of the target cell). This result indicates the advantage of convolutional-based models over conventional LUC models, which use only geographical features, when modeling the spatial pattern of LUC process.

\section{Discussion}

\subsection{Model visualization}

Although the explicit mechanism inside deep neural networks cannot be elucidated, some visualization techniques shed light on how conv-net and CDAE-net process the satellite images.
Figure 5 visualizes the activation maps from the first convolutional layers of conv-net and CDAE-net. 
The activation maps from conv-net and CDAE-net exhibit substantially different spatial patterns. This difference may be explained by the different purposes that the two models serve: conv-net seeks to capture the pattern that helps to explain the objective function, while CDAE-net seeks to capture the latent spatial features that help to explain the spatial variations in input images. Therefore, in conv-net, patterns with high activation values exhibit irregular shapes, but in CDAE-net, patterns with high activation values resemble the skeleton of geographical objects.

\begin{figure}
\centering
\includegraphics[width=17cm]{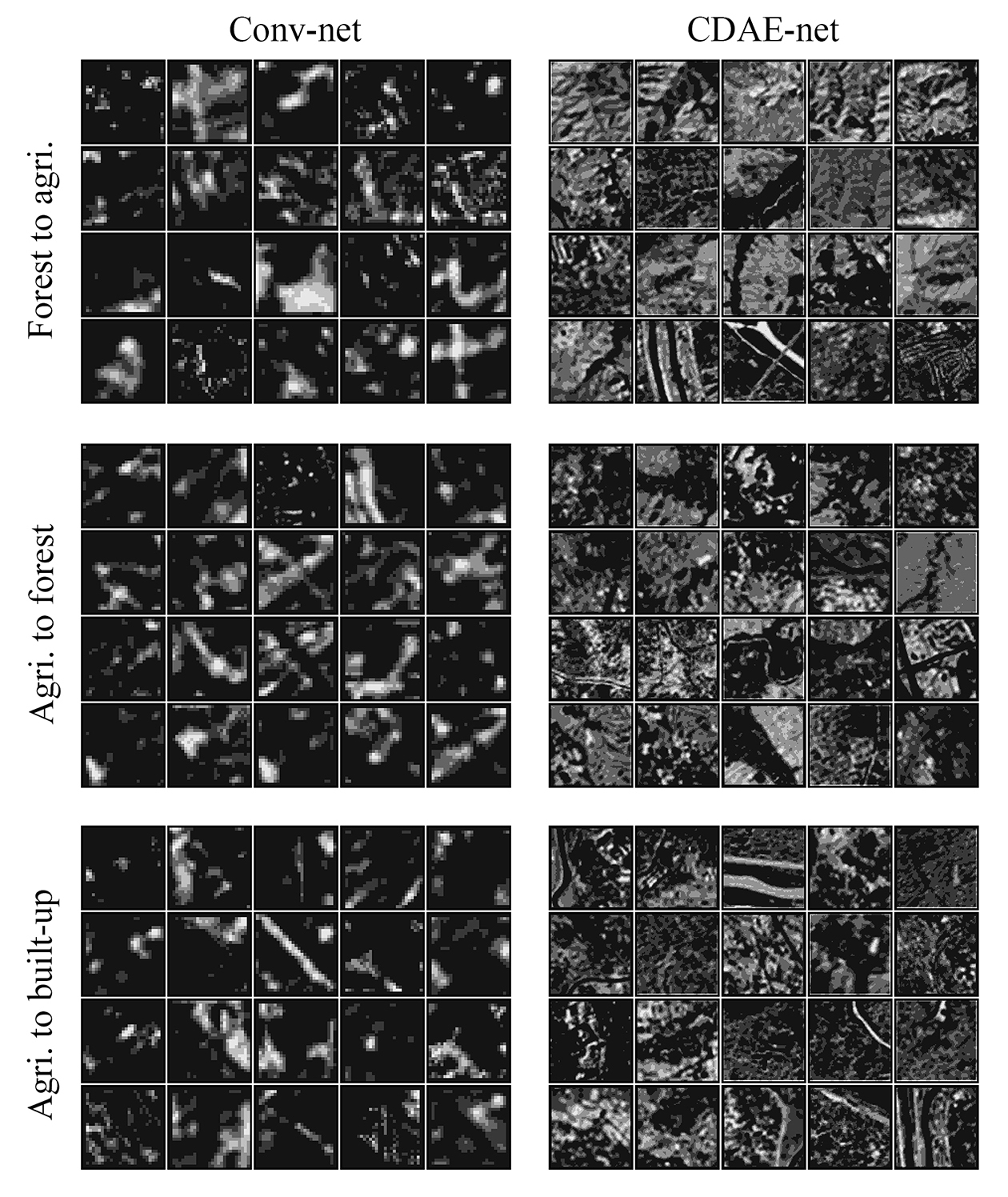}
\caption{The visualization of outputs from the first convolutional layers of the conv-nets and the CDAE-nets}
\end{figure}

We use t-distributed stochastic neighbor embedding (t-SNE) \citep{maaten2008visualizing} to visualize the distribution of spatial features extracted by the convolutional models. We randomly select a total of 2000 samples from the test set and feed them into conv-net and CDAE-net to obtain spatial features from the final convolutional layers. The spatial features are then embedded into 2D vectors by t-SNE. 
Figure 6 shows the results of t-SNE.
Theoretically, the distribution of spatial images in a 2D space may reflect the CNN effect, given that CNN would gradually transform the satellite image into linearly separable representations. 
As shown in Figure 6, the degree of sample aggregation with the same label is consistent with the accuracy of transition probability estimation; more samples with the same label aggregate together denote a higher predictive performance of the corresponding transition. For example, the transition from forest to agriculture has the highest predictive performance and the most visually separate spatial feature distribution. On the other hand, the transition from agriculture to built-up has the lowest predictive performance and barely separate spatial feature distribution.
Given that CDAE-net uses an unsupervised learning method, their embedded feature distributions are obviously different from those of conv-net. In terms of the transition from forest to agriculture, although the samples cluster into groups in both distributions, the clusters from CDAE-net appear to be more scattered than those from conv-net. 
To process the spatial features generated from the CDAE-net model, the classifier may require a higher level of non-linearity, which may partially explain the phenomenon in which the CDAE-net classifier generally has more hidden layers than conv-net's.

\begin{figure}
\centering
\includegraphics[width=17cm]{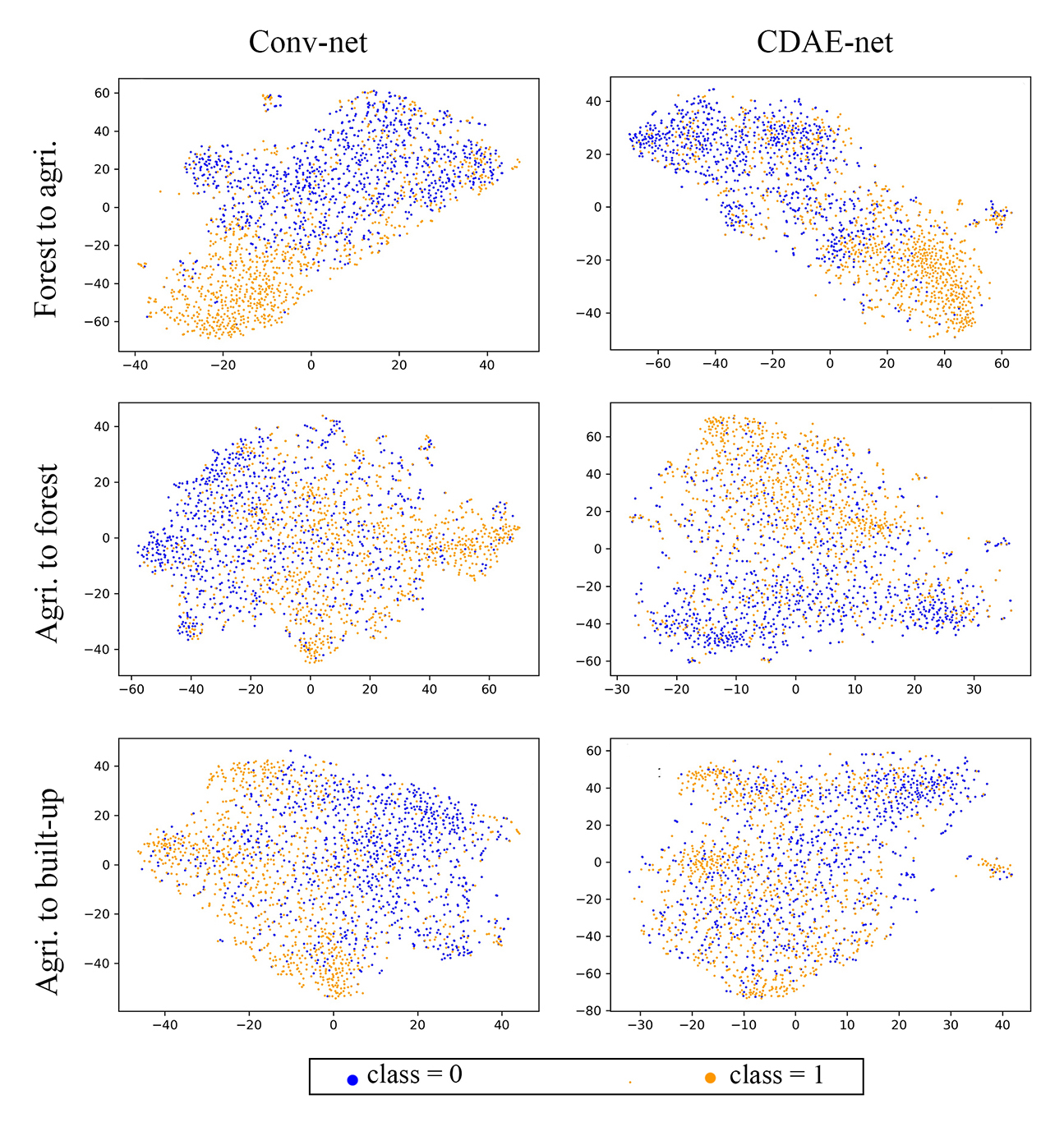}
\caption{The results of t-SNE for the spatial features that are extracted from satellite images by the conv-nets and CDAE-nets}
\end{figure}

\subsection{Model architecture}

We analyze the convolutional filter, the spatial weight layer and pooling within conv-net's architecture to examine its effect on the predictive performance of transition probability.
Table 7 presents the architectures of the baseline models for three transitions and their variants. 
For simplicity, given that the baseline models for three transitions have similar architectures as shown in Table 2, we use the same architecture for the three transitions in the analyses.
We also omit the classifier architectures, which are the same as those in Table 2.
All the models are independently developed and trained to facilitate an unbiased comparison. We find that the model's performance is sensitive to the weight and bias initialization; this trait causes some differences in the evaluation results compared with the evaluation results shown in Table 5.
Table 8 presents the corresponding results of analyses of varying filter size, spatial weight layer and pooling.

\begin{table}[h]
\small
\centering
\caption{The architectures of baseline models and their variants used for sensitivity analyses}
\label{my-label}
\begin{tabular}{| C{3cm} | C{3cm} | C{3cm} | C{3cm} | C{3cm} | }
\toprule
\multirow{2}{3cm}{Baseline models}       & \multicolumn{2}{| c |}{Variants: varying filter size}   & \multirow{2}{3cm}{Variant: no spatial weight} & \multirow{2}{3cm}{Variant: using average pooling} \\
\hhline{~--~~}
                       & $5\times5$                        & $7\times7$             &                   &                               \\
\midrule
conv-128               & conv-128                          & conv-128               & conv-128                                    & conv-128                      \\
spatial weight         & spatial weight                    & spatial weight         & max pooling                                 & spatial weight                \\
max pooling            & max pooling                       & max pooling            &                                             & average pooling               \\
\midrule
conv-256               & conv-256                          & conv-256               & conv-256                                    & conv-256                      \\
spatial weight         & spatial weight                    & spatial weight         & max pooling                                 & spatial weight                \\
max pooling            & max pooling                       & max pooling            &                                             & average pooling               \\
\midrule
conv-512               & conv-512                          & conv-512               & conv-512                                    & conv-512                      \\
spatial weight         & spatial weight                    & spatial weight         &                                             & spatial weight                \\
\midrule
conv-1024              & conv-1024                         & conv-1024              & conv-1024                                   & conv-1024                     \\
spatial weight         & spatial weight                    & spatial weight         &                                             & spatial weight                \\
\midrule
\multicolumn{5}{| c |}{global average pooling}         \\
\bottomrule
\end{tabular}
\footnotesize\flushleft{
\textit{Notes:}\\
1. The same convolutional architecture is used for the LUC models of all three transitions. \\
2. The architectures of classifiers are the same as those in Table 2 and thus are omitted. \\
3. Conv-N denotes a convolutional block composing of a convolutional layer with $3\times3$ filter size and N of filters, a Batchnorm layer and a ReLU layer. \\
4. Pooling layer has a kernel size of $3\times3$ and a stride of 3.
}
\end{table}

\begin{table}[h]
\small
\centering
\caption{Results of the sensitivity analyses with respect to filter size, spatial weight layer and pooling}
\begin{tabular}{p{5cm} lccc}
\toprule
(a) Varying filter size               &               &                 &                 &                   \\
\midrule
                                       &               & Forest to agri. & Agri. to forest & Agri. to built-up \\
\hhline{~~---}
\multirow{2}{5cm}{Baseline ($3\times3$)}                  & AUC-ROC       & 0.941           & 0.904           & 0.694             \\
                                       & AUC-PR        & 0.711           & 0.478           & 0.238             \\
\multirow{2}{5cm}{Variant ($5\times5$)}                   & AUC-ROC       & 0.923           & 0.897           & 0.671             \\
                                       & AUC-PR        & 0.698           & 0.447           & 0.223             \\
\multirow{2}{5cm}{Variant ($7\times7$)}                   & AUC-ROC       & 0.901           & 0.860            & 0.648             \\
                                       & AUC-PR        & 0.625           & 0.419           & 0.209             \\
\midrule
(b) Spatial weight layer               &               &                 &                 &                   \\
\midrule
                                       &               & Forest to agri. & Agri. to forest & Agri. to built-up \\
\hhline{~~---}
\multirow{2}{5cm}{Baseline (with spatial weight)}          & AUC-ROC       & 0.939           & 0.906           & 0.693             \\
                                       & AUC-PR        & 0.712           & 0.470           & 0.239             \\
\multirow{2}{5cm}{Variant (no spatial weight)}            & AUC-ROC       & 0.924           & 0.906           & 0.684             \\
                                       & AUC-PR        & 0.688           & 0.447           & 0.232              \\
\midrule
( c ) Max v.s. average pooling &               &                 &                 &                   \\
\midrule
                                       &               & Forest to agri. & Agri. to forest & Agri. to built-up \\
\hhline{~~---}
\multirow{2}{5cm}{Baseline (max pooling)}                 & Training loss & 0.317           & 0.522           & 0.423             \\
                                       & Test loss     & 0.388           & 0.603           & 0.512             \\
\multirow{2}{5cm}{Variant (average pooling)}              & Training loss & 0.295           & 0.508           & 0.412             \\
                                       & Test loss     & 0.410           & 0.624           & 0.527            \\
\bottomrule
\end{tabular}
\footnotesize\flushleft{
\textit{Notes:}\\
Training loss and test loss are the binary cross entropy losses of training set and test set, respectively.
}
\end{table}

The results comparison on the varying filter sizes show that the predictive performances decrease significantly as the filter size increases. In terms of the transition from forest to agriculture, the AUC-PR decreases by approximately 11\% from 0.69 to 0.61, indicating that conv-net's predictive performance is sensitive to the filter size.
Furthermore, the improved performance gained by the smaller filter size implies that the large filter size is not necessary to address the redundancy of spatial information in the satellite images.

The spatial weight layer analysis is designed to identify the benefit of incorporating the specific regularization on the spatial features. 
The results show that AUC-ROC and AUC-PR gradually increase by approximately 0.3 from the variant model without the spatial weight layer on the baseline models; this result indicates the effectiveness of the spatial weight layer in improving the predictive performance. However, the spatial weight layer is specifically designed as a regularization method for spatial features extracted from the satellite images, and its effect depends on the representative degree of the spatial features; for example, the performance improvement for the transition from agriculture to built-up is much smaller than forest to agriculture.

To examine the pooling effect, we present the binary cross-entropy losses of the training and test sets instead of the AUC to reflect the model capacity and generalization performances. 
The results show that the models using max pooling exhibit higher training and lower test losses than the models using average pooling; models using max pooling have smaller capacity but better generalization ability, while models using average pooling have larger capacity but limited generalization ability. 
If pooling is considered as a prior on the spatial features, max pooling imposes a stronger prior than average pooling. This result is because max pooling replaces the values within a kernel with their maximum value rather than with the average. Hence, a stronger prior may be more beneficial for filtering out the useful features from the satellite images.

\section{Concluding remarks}

This study applies CNN to enhance the performance of LUC modeling. We developed two convolutional-based models, conv-net and CDAE-net, to estimate three types of LU transition probabilities: forest to agriculture, agriculture to built-up and agriculture to forest. The results show that both conv-net and CDAE-net improve the accuracy of transition probability estimation compared with the MLP estimator, which has conventional geographical features as its sole input. Moreover, conv-net and CDAE-net achieve similar predictive performances of the estimation transition probabilities between forest and agriculture. On the other hand, CDAE-net significantly outperforms conv-net when estimating transition probability from agriculture to built-up. This result may be explained by CDAE-net's relatively more effective task handling performance for relatively complicated transitional rules and/or data with higher noise because it can learn latent representations and the denoising design.

This study's results provide several useful findings on convolutional-based model architecture.
1) Shallow architecture is sufficient for the LUC modeling task in this study. Conv-net and CDAE-net have only four and two convolutional layers, respectively; the layers are rather shallow compared with commonly used layers in computer vision studies.
2) The LUC models learn different transitional rules per the LUC process, and the model architectures could vary. Although the classifier architectures are different, their convolutional architectures are very similar for the transitions considered in this analysis. This observation indicates that the spatial features are extracted using similar learning processes. Hence, the extracted spatial features from the satellite images have similar degrees of complexity. 
3) The spatial weight layer, which is specifically designed to apply distance-decay regularization on spatial features, effectively improves conv-net's predictive performance.

The approach developed in this study can be further adapted to broader applications. Due to the data limitation, we used GLS satellite dataset based on Landsat with relatively low resolution but the future studies may apply the models to finer satellite imagery such as SPOT. The rich spatial information may further improve the LUC modeling performance. 
Furthermore, we recommend to combine convolutional-based models with classifier other than the neural networks. For example, the CDAE model may be combined with ensemble models such as random forest to further incorporate stochastic process for transitional rule determination.

\bibliographystyle{elsarticle-harv} 
\bibliography{ref}

\end{document}